\documentclass[a4paper, preprint, superscriptaddress]{revtex4-1}

\usepackage{hyperref}
\usepackage{graphicx}
\usepackage{amsmath}
\usepackage{wrapfig}
\usepackage{url}
\usepackage{xcolor}
\usepackage{upgreek}
\usepackage{subfigure}
\usepackage{gensymb}

\newcommand{\boutxx}{\texttt{BOUT++ }}
\newcommand{\hermes}{\texttt{Hermes }}

\begin{document}

\title{Global fluid turbulence simulations in the SOL of a stellarator island divertor}

\author{B Shanahan}
\email{brendan.shanahan@ipp.mpg.de}
\author{D Bold}

\affiliation{Max-Planck-Institut f\"ur Plasmaphysik, Wendelsteinstr. 1, 17491 Greifswald, Deutschland}
\author{B Dudson}
\affiliation{Lawrence Livermore National Laboratory, 7000 East Avenue, Livermore, 94550, CA, USA}




\begin{abstract}
Isothermal fluid turbulence simulations have been performed in the edge and scrape-off-layer (SOL) of an analytic stellarator configuration with an island divertor, thereby providing numerical insight into edge turbulence in regions around islands in a stellarator. The steady-state transport follows the 1/R curvature drive toward the outboard side, but large fluctuations are present throughout the island divertor region, with the average wavelength of similar size to the island width. The system exhibits a prominent m=2, n=5 mode, although other modes are present. The amplitude and radial extent of the density fluctuations are similar throughout the edge and SOL, but can decrease near island O-points. The fluctuations exhibit a predominantly positive skewness on the outboard midplane, indicating blob-like perturbations for the transport into the outer SOL. It is determined that a point on the separatrix is generally more correlated with regions outside of the SOL than a nearby reference point which does not lie on the separatrix.
\end{abstract}
\maketitle

\section{Introduction}
Advances in stellarator optimization have led to unprecedented improvement in neoclassical transport in Wendelstein 7-X~\citep{Beidler2021} such that anomalous transport is now contributing a significant portion of the transport, especially in the Scrape-Off-Layer (SOL)~\citep{Pablant2018}. The island divertor concept, originally proposed nearly five decades ago~\citep{Karger1977}, utilizes low-order magnetic islands which intersect the wall to manage heat and particle exhaust~\citep{Feng2006, Feng2022}. The island divertor is advantageous in that it potentially provides a higher whetted area, and has been able to enter very stable detachment regimes~\citep{Pedersen2019}. Due to this island divertor, the SOL of W7-X provides a unique environment for SOL physics, as the toroidally-discontinuous intersection of magnetic islands creates complicated topologies which can inhibit numerical investigation. As such, turbulence simulations in stellarators have focused on the core transport using either a gyrokinetic~\citep{Xanthopoulos2007, Maurer2020, BañónNavarro2020, Singh2022} or, less commonly, a fluid approach~\citep{Kleiber2005}. Recent work using local fluid simulations has informed interpretation of experimental measurements~\citep{ShanahanJPCS2018, KillerShanahan2020, Shanahan2021, Huslage2024}, without requiring modelling of the complex and numerically challenging SOL topology. Developments of numerical methods have provided the opportunity for global fluid turbulence simulations in stellarator geometries~\citep{Shanahan2018, Coelho2022}, but a general understanding of global phenomena at the edge and SOL of stellarators is only in its infancy. In this work, we utilize an isothermal model in the~\boutxx~\citep{Dudson2009} framework to explore the nature of turbulence in the island divertor region at the edge of an analytic stellarator configuration~\citep{Coelho2022}.
 
Section~\ref{sec:methods} details the methods used in this work, including the plasma model, geometry, and computational grid. Section~\ref{sec:results} discusses the simulation results, and is divided into two subsections; the steady-state transport properties of the system and the dynamics of the fluctuations in the island SOL. The implications of this work and its context within previous work are discussed in Section~\ref{sec:discussion}.

\section{Methods}
\label{sec:methods}
The strength of~\boutxx~\citep{Dudson2009} is its flexibility: models and numerical methods can be easily changed to suit the problem to be addressed.  Here, we use an isothermal plasma model in a complex numerical scheme, as the turbulent dynamics are available despite numerous assumptions, but the geometry is necessarily complicated. 
\subsection{Isothermal plasma model}
In this work we exploit a reduced MHD model from the \hermes family of models~\citep{Dudson2017, Leddy2016, Huslage2024}, which do not separate fluctuations from the background. The simulations here are isothermal, and evolve number density $n$, vorticity $\omega$, parallel ion momentum $m_inv_\parallel$ and Ohm's law $J_\parallel = en(v_{\parallel,i} - v_{\parallel,e}) = -\frac{1}{\nu} \partial_\parallel \phi - \frac{1}{n_e} \partial_\parallel p_e$ for the parallel electron velocity $v_{||e}$. Quasineutrality is assumed, so that the electron and ion  densities are equal: $n_e = n_i = n$. Here, $\nu=1.96 \tau_{ei} \frac{m_i}{m_e}$  is the resistivity. The thin layer (Oberbeck-Boussinesq~\citep{Oberbeck1879}) approximation is made in calculating the potential~$\phi$ from vorticity $\omega$ such that:
\begin{equation}
\omega = \nabla\cdot\left[\frac{en_0}{\Omega B}\nabla_\perp\phi\right]
\end{equation}
where $\Omega = eB/m_i$ is the ion cyclotron frequency, and $n_0$ is a constant typical number density, in this work set to be $n_0 = 1\times10^{18}\mathrm{m}^{-3}$. The resulting equations for (electron) number density $n$, vorticity $\omega$ and parallel momentum $m_in v_{\parallel i}$ are:
\begin{eqnarray}
  %
  %
  \frac{\partial n}{\partial t} &=& -\nabla\cdot\left(n\mathbf{V}_{E\times B} + n\mathbf{V}_{mag,e}\right) - \nabla_{||}\left(n v_{||e}\right) + S_n 
\end{eqnarray}
\begin{eqnarray}
%
%
  \frac{\partial \omega}{\partial t} &=& \nabla\cdot\left[e\left(p_e + p_i\right)\nabla\times\frac{\mathbf{b}}{B}\right] \label{eq:vort_diamag} + \nabla_{||}j_{||} \label{eq:vort_jpar} - \nabla \cdot \left(\omega\mathbf{V}_{E\times B}\right)  \label{eq:vort_jpol}
\end{eqnarray}
\begin{eqnarray}
%
%
  \frac{\partial}{\partial t}\left(m_inv_{||i}\right) &=& -\nabla\cdot\left[m_inv_{||i}\left(\mathbf{V}_{E\times B} + \mathbf{b}v_{||i} + \mathbf{V}_{mag,i}\right)\right] - \partial_{||}p_e - \partial_{||}p_i
\end{eqnarray}
where $S_n$ is the density source, $p_{e/i}$ is the pressure for electrons and ions respectively, and the drift terms for $E\times B$, $\mathbf{V}_{E\times B}$, and ion and electron magnetic drifts, $\mathbf{V}_{mag,e/i}$, are defined as:
\begin{equation}
\mathbf{V}_{E\times B} = \frac{\mathbf{b}\times\nabla \phi}{B} \qquad
\mathbf{V}_{mag,e} = -\frac{T_e}{e}\nabla\times\frac{\mathbf{b}}{B} \qquad \mathbf{V}_{mag,i} = \frac{T_i}{e}\nabla\times\frac{\mathbf{b}}{B}
\end{equation}
Here the notation is such that $\nabla_\perp = \nabla - \bf{bb}\cdot \nabla$, $\partial_\parallel f = {\bf{b}}\cdot\nabla f$, and $\nabla_\parallel f = \nabla \cdot ({\bf{b}}f)$.  While the model allows terms for anomalous diffusion, these terms are not necessary for numerical stability and, as the nature of their origin is unknown, are neglected in this work. An imposed parallel diffusion of 4.2$\mathrm{m^2/s}$ is used in Ohm's law to suppress numerical instability.

\begin{figure}
  \centering
  \includegraphics[width=1\columnwidth]{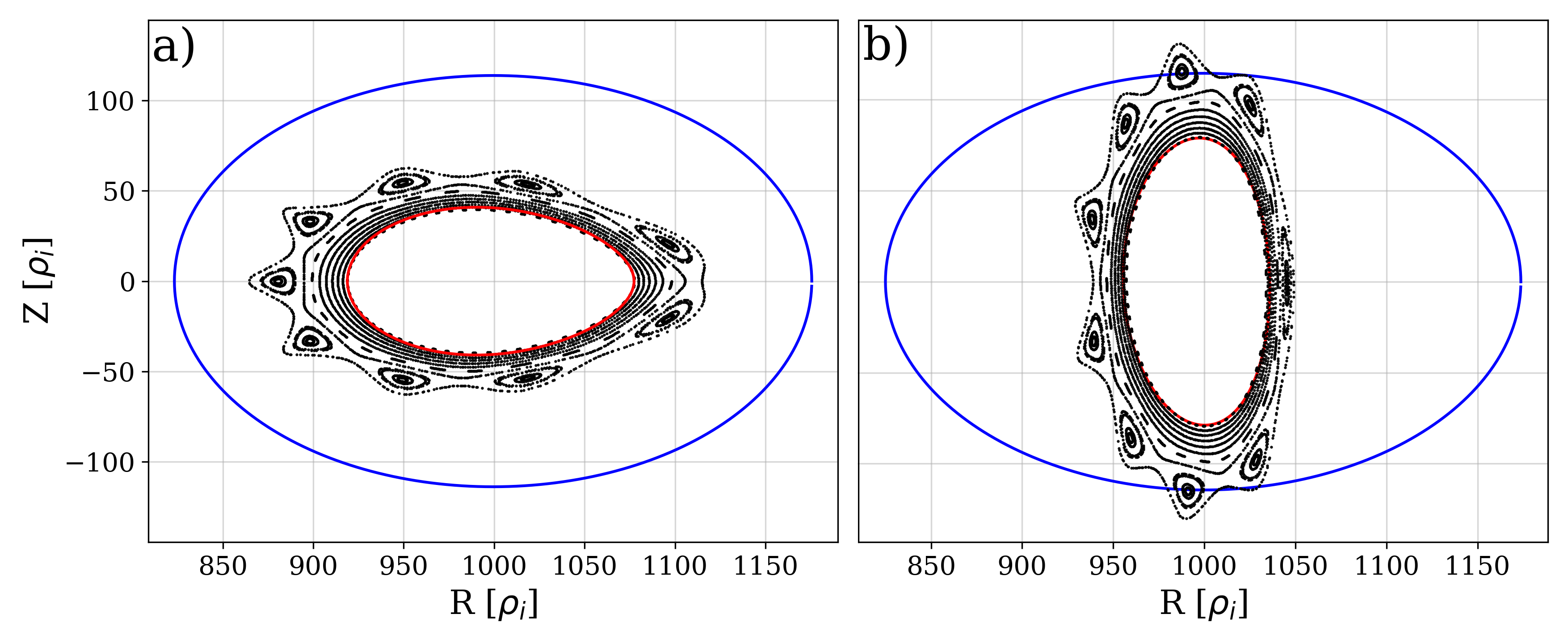}
  \caption[]{Poincar\'e plots (black) of the magnetic field using parameters from~\citep{Coelho2022}, including the inner (red) and outer (blue) surfaces of the grid.}
  \label{fig:poincare}
\end{figure}

\subsection{BSTING and \boutxx}

The BSTING project~\citep{Shanahan2018} has recently allowed for simulations of stellarator geometries using the~\boutxx framework~\citep{Dudson2009}, in which exploitation of the Flux-Coordinate-Independent (FCI) method for parallel derivatives~\citep{Hariri2013,Hariri2014,Hill2015,Hill2017,ShanahanJPCS2016, Stegmeir2016} allows for simulation of complex magnetic geometries including magnetic islands and chaotic field regions.  The implementation of 3D metric tensor components in~\boutxx has allowed for poloidally-curvilinear FCI grids~\citep{Shanahan2018} that do not include grid points in the core of the plasma -- where the fluid approximation due to high collisionality can break down -- and allow for the poloidal grid alignment to flux surfaces or plasma-facing-components.  As such, FCI simulations can be performed with resolutions comparable to field-aligned tokamak simulations. 

\subsection{Geometry and initial conditions}
The simulation geometry used here is similar to that in~\citep{Coelho2022}.  Namely, a stellarator magnetic field with an outer $m=9$ island chain was created using the same Dommaschk potential~\citep{Dommaschk1986} as in~\citep{Coelho2022}, where the magnetic field varies to lowest order by $1/R$ -- similar to (for instance) a tokamak.  The BSTING framework utilizes curvilinear grids, however, meaning that in contrast to the work presented in~\citep{Coelho2022}, here the core is neglected, and the outer poloidal surface is an ellipse with the same maximum dimensions of the outer boundary used in~\citep{Coelho2022}. The inner surface is aligned to a vacuum flux surface, and the elliptical outer surface provides a plasma facing component which intersects islands at discrete toroidal locations (for instance, Figure~\ref{fig:poincare}b), as in an island divertor.

The numerical grid utilized in this work was of the size (radial, toroidal, poloidal) $(nx,ny,nz) = (68,128,256)$ with an average resolution of $(dx,dy,dz) \approx (1,75,1)\rho_i$, where $\rho_i$ is the ion Larmor radius. The simulation domain spans the entire toroidal extent ($0$ to $2\pi$). Despite the difference in numerical implementation, the magnetic field is identical to that in~\citep{Coelho2022}, including the outer magnetic islands, as shown in Figure~\ref{fig:poincare}. The radial extent of the SOL can be as small as 10 Larmor radii, and perturbations often exhibit a similar perpendicular size. As such, fluctuations cannot be categorized as edge- or SOL-localized. A more detailed discussion of the fluctuations follows in Section~\ref{sec:perturbations}.

The simulations are flux-driven, with a particle source of the form:
\begin{equation}
    S_n(r) = S_0\frac{1}{w\sqrt{2\pi}} exp\left(\frac{-(r-r_0)^2}{(2w^2)}\right) \mathrm{m^{-3}{s^-1}}
\end{equation}  
where $S_0$ is a prescribed constant used to balance the sinks, $r_0$ is the radial location the peak, and $w$ is the Gaussian RMS width. Here, $r_0$ and $w$ are chosen such that the source peaks at the same radial location in the closed-field-line region as the source in~\citep{Coelho2022}. Since the particle source introduces an external drive into the system, dynamics which originate farther inward (toward the core) must be carefully treated -- and are therefore not included in the computational domain.  The equilibrium parameters -- the background density ($n_0 = \mathrm{1\times10^{18}m^{-3}}$), the magnetic field ($B_0 \approx 0.3T$) and the background temperature ($T_0 = 10eV$) have been chosen to most closely align with~\citep{Coelho2022}. These parameters are similar to those found in university-scale experiments, such as TORPEX~\citep{Furno2008, Shanahan2016}. Parallel sheath boundary conditions~\citep{Walkden2016} are imposed such that the species are accelerated to the local sound speed and implemented using the Leg-Boundary-Fill method at the ends of magnetic field lines~\citep{Hill2017}.  Perpendicular boundary conditions, with the exception of plasma potential, are zero-gradient allowing flows into the domain from the inner boundary and out of the domain at the outer boundary. The plasma potential is calculated from vorticity using PETSc~\citep{Balay1997} with Dirichlet boundary conditions such that $\phi_{boundary} = 2.8T_e$. The density is set to an approximate initial profile, and an initial perturbation in vorticity is added to the system; after a transient phase, the simulation settles into a state where the particle source is balanced by the sinks in the sheath, characterized by fluctuations throughout the domain. The simulation is then run such that the timescale exceeds 1ms, which is calculated in approximately 30,000 core-hours. 

\section{Results}
The dynamics simulated here can be separated into two scales; large-scale, steady-state flows of the plasma and the perturbations on top of these macroscopic aspects.  The following two subsections will look into these dynamics in more detail.
\label{sec:results}
\subsection{Macroscopic transport and steady-state flows}
\begin{figure} 
  \centering
  \includegraphics[width=1\columnwidth]{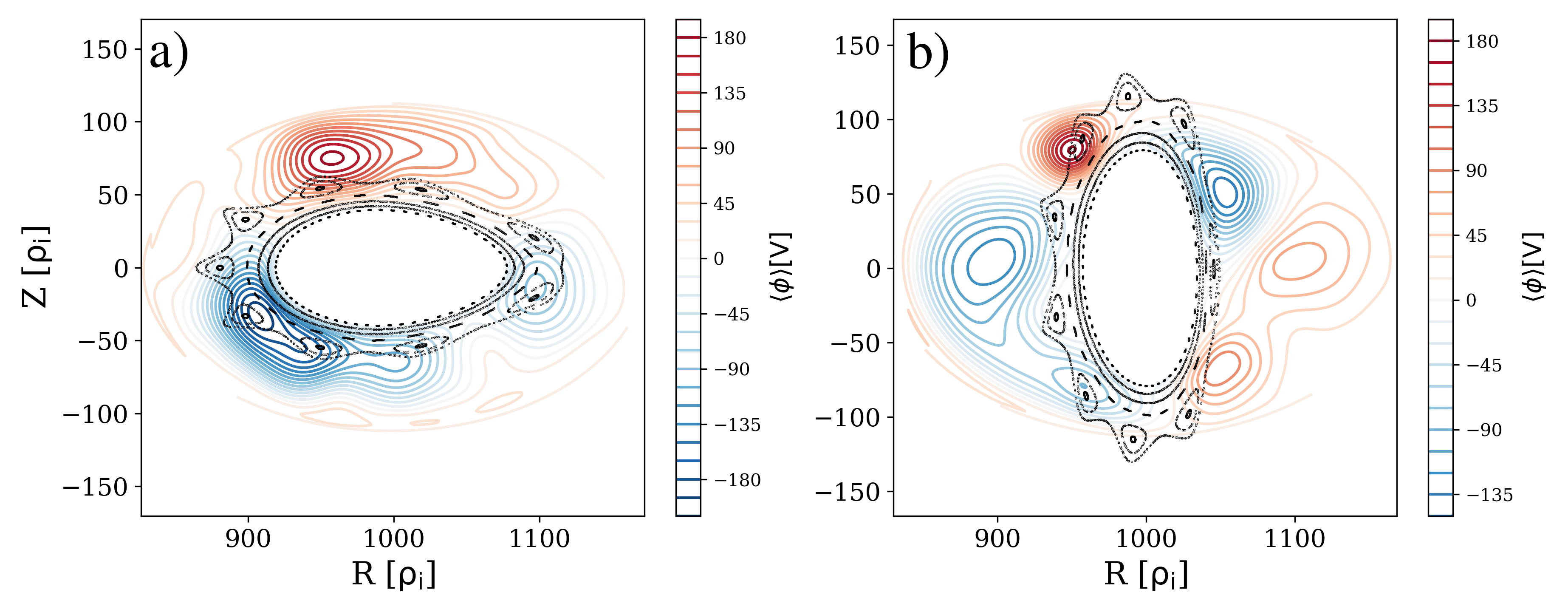}
  \caption[]{Time-averaged plasma potential; the contour lines indicate flows due to $E\times B$ advection. Transport towards the outboard side due to 1/R curvature drive is the dominant flow within the system. In this figure, the magnetic field is predominantly into the page.}
  \label{fig:phimean}
\end{figure}

\begin{figure}
  \centering
  \includegraphics[width=1\columnwidth]{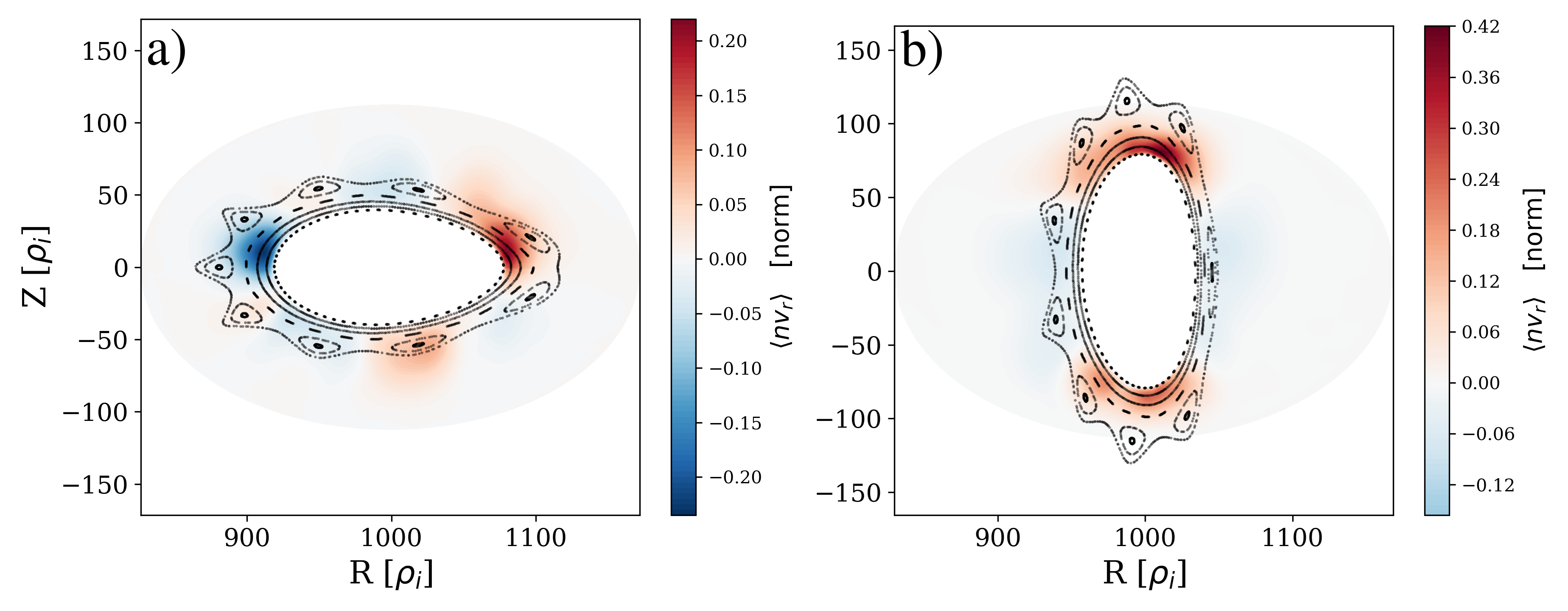}
  \caption[]{Time-averaged radial flux, indicating regions of radial transport. A negative radial flux indicates radially-inward flux, whereas a positive flux is outward.}
  \label{fig:nvr}
\end{figure}

On longer time scales ($t > 100$\,µs), the plasma advects with a dominant motion toward the outboard side. This motion is visualized by plotting the contours of the time-averaged plasma potential, Figure~\ref{fig:phimean}, as the flows follow the contours of $\left<\phi\right>$ via $E\times B$ motion. Throughout this work, the angled brackets ($\left<\right>$) will always indicate a time average, and tilde ( $\tilde{}$ ) will indicate a perturbed quantity around the time average.
The time-averaged transport indicates predominantly radial transport in the edge and SOL, with a return flow antiparallel to the curvature drive in the low-density region -- near the outer boundary -- best visualized in Figure~\ref{fig:phimean}a. The dynamics are more complicated in the vertical cross section, Figure~\ref{fig:phimean}b, due to the sheath connection that introduces a discontinuity between the inboard and outboard. A return flow is seen on the outboard midplane of the vertical cross section, and can also be seen when plotting the radial flux~\citep{Shanahan2014}, $\Gamma_r = \left<nv_r\right>$, as is shown in Figure~\ref{fig:nvr}. 
The most prominent areas of radial flux coincide with the strongest gradients in potential contours shown in Figure~\ref{fig:phimean} -- for instance on the inboard side of Figure~\ref{fig:phimean}a. These areas of large radial flux can be attributed to steady-state flows due to the 1/R curvature drive; a radially-inward flux is seen on the inboard side of Figure~\ref{fig:nvr}a in blue, and a strong radially-outward flux on the outboard side in red. The vertical cross section, Figure~\ref{fig:nvr}b, diverges from this curvature-driven nature, with flux on the inboard side prominent near X-points -- including a flux antiparallel to the curvature drive at the midplane -- and significant outward radial flux near the intersection with the outer boundary.

\begin{figure}
  \centering
  \includegraphics[width=1\columnwidth]{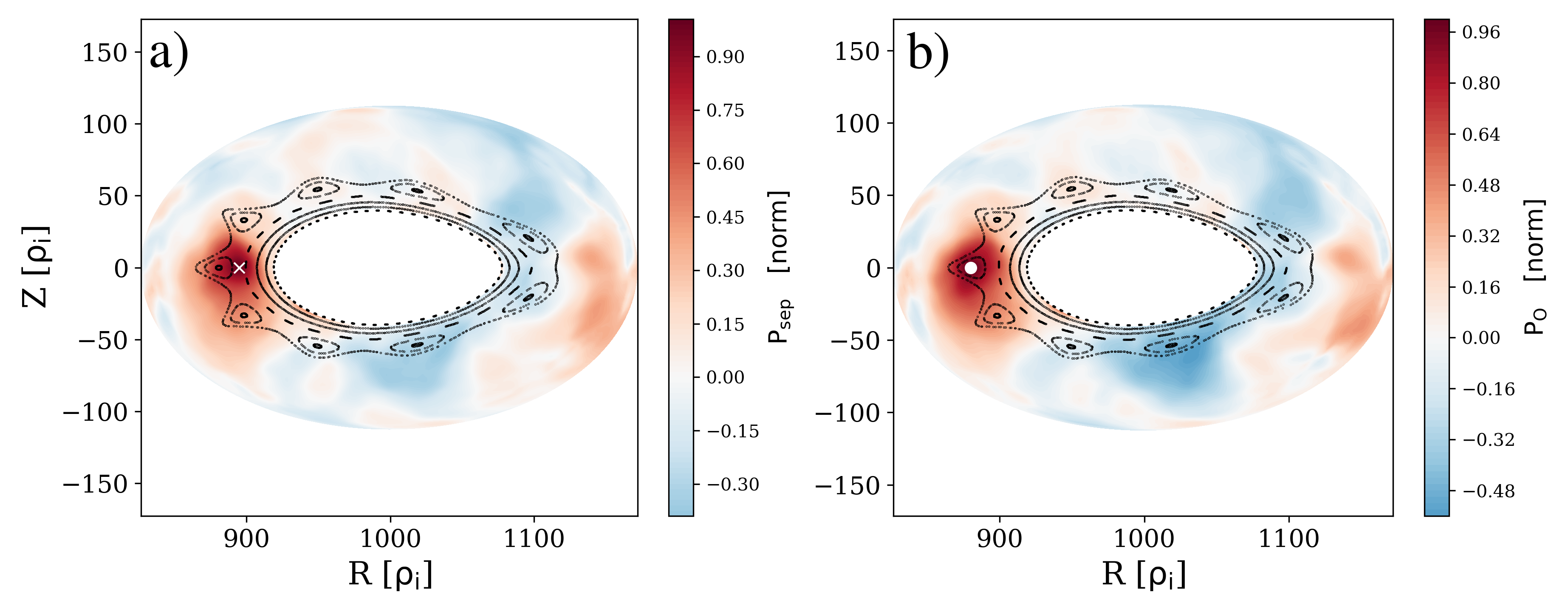}
  \caption[]{The Pearson correlation coefficients when considering a reference point a) in the separatrix, $P_{sep}$, and b) when choosing a reference point at a magnetic O-point, $P_O$. These reference points are marked with a white X and a white dot, respectively. A positive value indicates positive correlation with the reference point.}
  \label{fig:cross_corr_comp}
\end{figure}

To further examine the flows in the island SOL presented here, one can plot the Pearson correlation coefficient~\citep{Freedman2007} of density signals relative to a given point in the SOL, as is done in~\citep{Shanahan2014}. In this work, we  correlate the time trace of every point in the domain to a reference point, defining the Pearson correlation coefficient $P_{fg}$ between a reference point $f$ and a sample point $g$ as:
\begin{equation}\label{eq:xcorr}
    P_{fg} = \frac{\mathrm{cov}(f,g)}{\sigma_f \sigma_g} = \sum^{N_t-1}_{t=0}\left(f_t - \left<f\right>\right)\left(g_t - \left<g\right>\right) \left\{\left[\sum^{N_t-1}_{t=0}\left(f_t - \left<f\right>\right)^2\right]\left[\sum^{N_t-1}_{t=0}\left(g_t - \left<g\right>\right)^2\right]\right\}^{-\frac{1}{2}}
\end{equation}
where $\mathrm{cov}(f,g)$ indicates the covariance of $f$ and $g$, $\sigma_f$ and $\sigma_g$ are the standard deviations of $f$ and $g$ respectively, the angled brackets ($\left<\right>$) indicate a time average, $t$ is the time index for each time trace and $N_t$ is the number of time points in the time trace. 
A higher correlation of a point in space to a reference point indicates a better correlation between fluctuations at these positions. Due to the large-scale fluctuations, the correlation $P_{fg}$ often has a similar structure regardless of the reference point chosen, see Figure~\ref{fig:cross_corr_comp}.

The Pearson correlations shown in Figure~\ref{fig:cross_corr_comp} are very similar in structure, inhibiting a clear conclusion for the role of the separatrix in providing a channel for flows in the SOL. For this reason, we will look at the relative correlation when choosing two reference points; one correlation with a reference point at the separatrix $P_{sep}$ , and one with a reference point at an island O-point, $P_O$. 

\begin{figure}
  \centering
  \includegraphics[width=1\columnwidth]{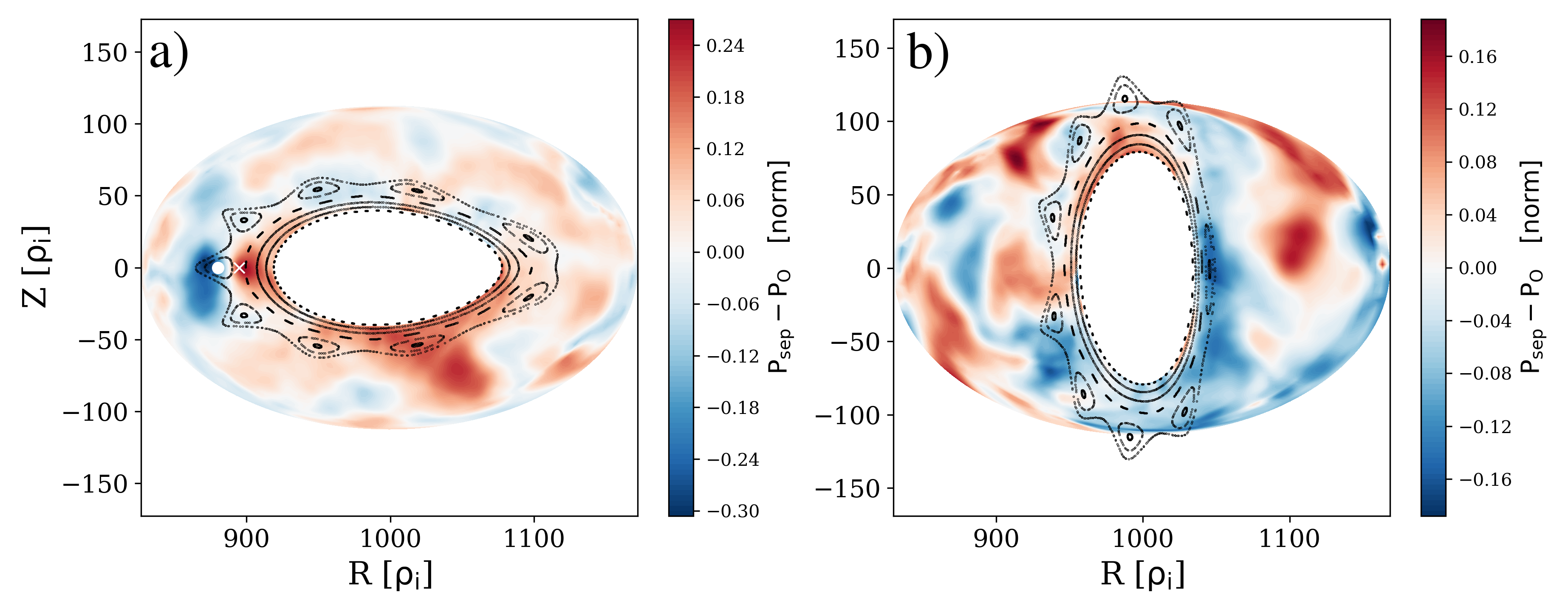}
  \caption[]{The relative Pearson correlation coefficients when considering a reference point in the separatrix, $P_{sep}$, and when choosing a reference point at a magnetic O-point, $P_O$. These reference points are marked with a white X and and circle in Figure~\ref{fig:cross_corr_diff}a. A positive value indicates stronger correlations with fluctuations at the separatrix, whereas a negative value indicates correlation with fluctuations at an island O-point.}
  \label{fig:cross_corr_diff}
\end{figure}

Figure~\ref{fig:cross_corr_diff} shows the difference of two correlations given by Equation~\ref{eq:xcorr}; one where the reference point is in the separatrix, $f_{sep}$, and one where the reference point is at an O-point, $f_O$. The reference points are shown as white markers in Figure~\ref{fig:cross_corr_diff}a.  By plotting their difference, $P_{sep} - P_O$, where the higher correlation of the separatrix to the outer edges of the domain is illustrated by the positive difference of the two correlations, one can obtain a clearer picture of the correlation within the SOL.
Figure~\ref{fig:cross_corr_diff}a indicates an increased relative correlation of the edge SOL to the separatrix -- even on the outboard side -- while the strongest relative correlation to the O-point is seen nearest the O-point reference location. The vertical cross section, Figure~\ref{fig:cross_corr_diff}b, does not indicate such strong correlation with the separatrix reference location, except at the edges near the boundary. The higher relative correlations of the boundary on the vertical cross section indicate that the separatrix could provide a transport channel to the boundary.

\subsection{Perturbations near the edge and SOL}
\label{sec:perturbations}
\begin{figure}
  \centering
  \includegraphics[width=1\columnwidth]{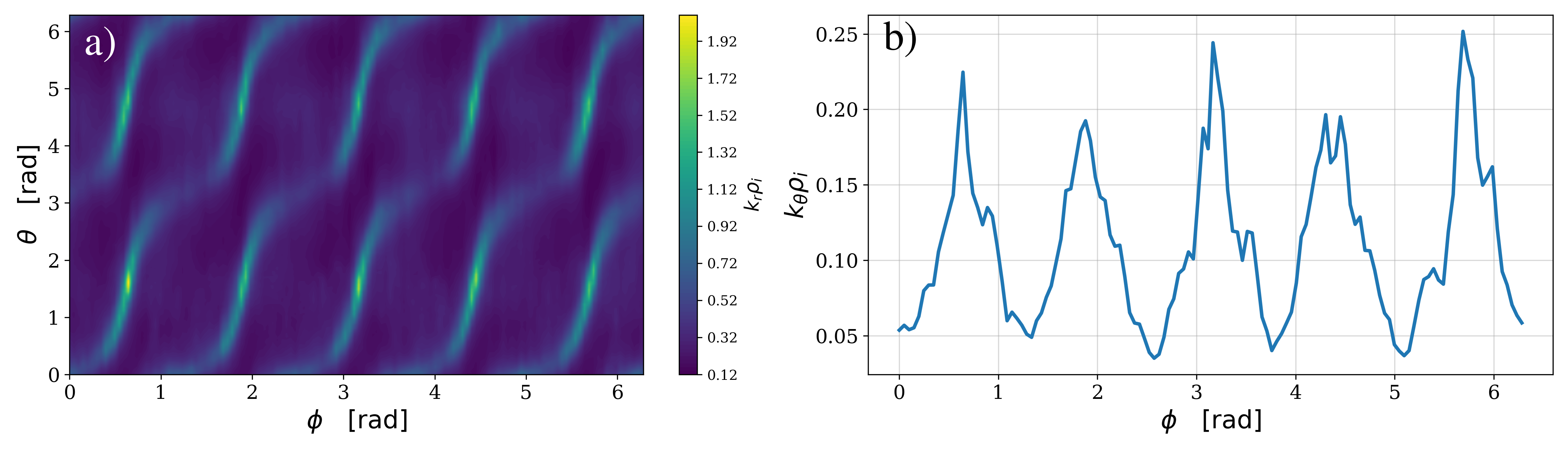}
  \caption[]{a) average radial fluctuation size and b) average poloidal fluctuation size on a flux surface indicated by the dashed surface in Figure~\ref{fig:poincare}.}
  \label{fig:kperp}
\end{figure}

Having examined the characteristics of the large-scale dynamics, attention is now turned to the dynamics of the perturbations. The system exhibits a strong m=2, n=5 mode, although other modes are present. The fluctuations are large -- $k_\perp \rho_s$ is generally less than 1, as indicated in Figure~\ref{fig:kperp}. As the magnetic islands are, at widest, only a couple 10s of $\rho_i$ the fluctuations can at times exceed the width of the island SOL.

\begin{figure}
  \centering
  \includegraphics[width=1\columnwidth]{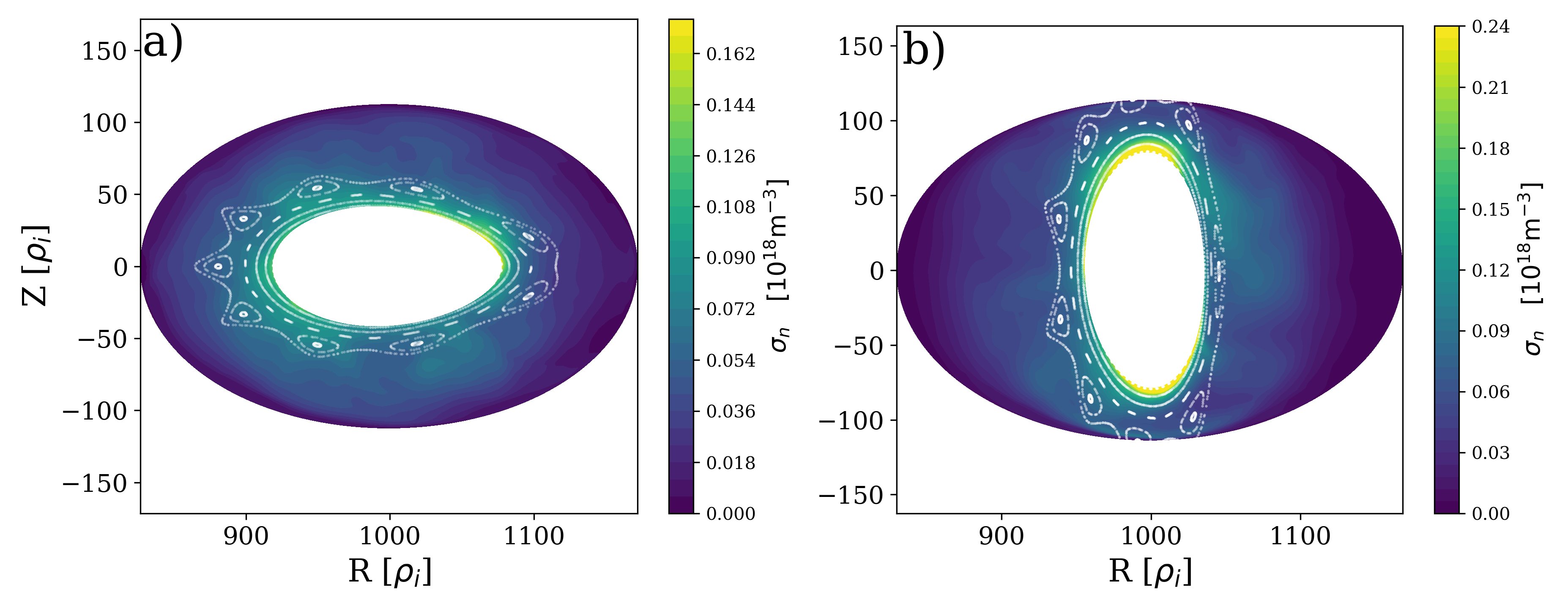}
  \caption[]{Standard deviation of the density; perturbations are apparent at all poloidal locations, with a slight increase on the outboard side.}
  \label{fig:sigma_n}
\end{figure}

\begin{figure}
  \centering
  \includegraphics[width=1\columnwidth]{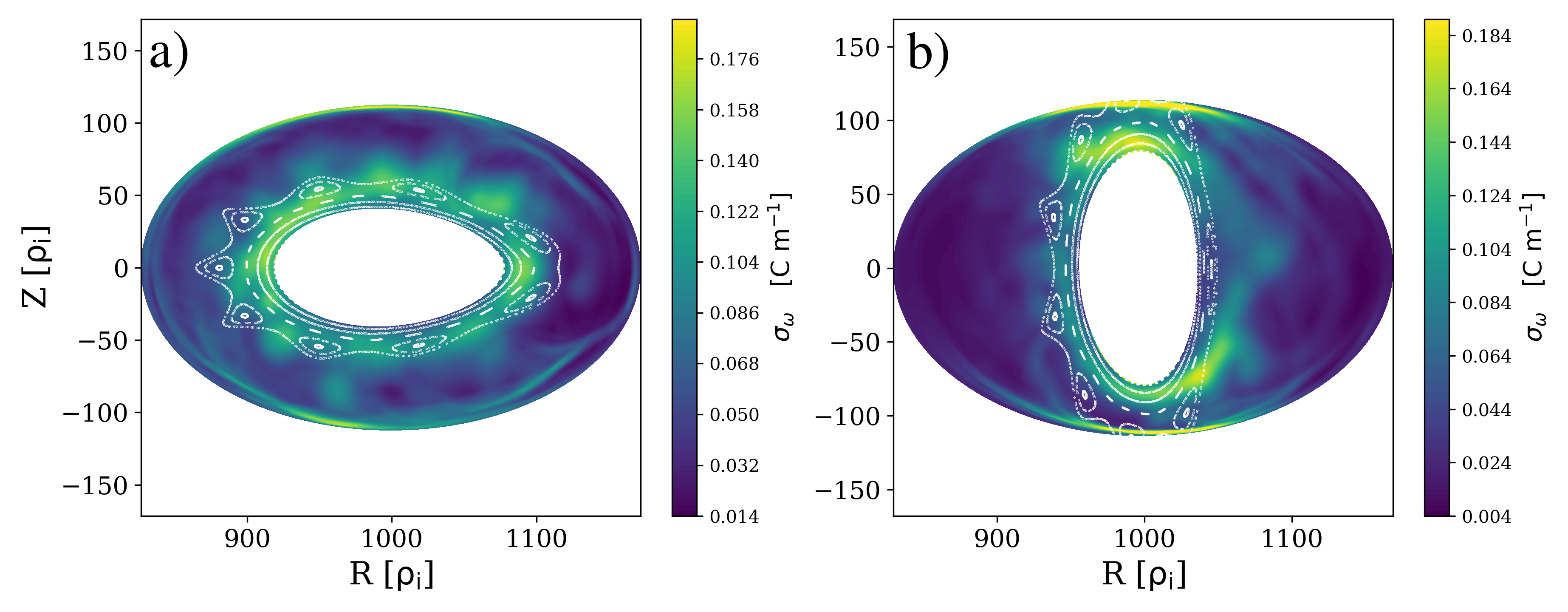}
  \caption[]{Standard deviation of vorticity. The horizontal cross section, subfigure a, indicates better poloidal symmetry compared to the vertical cross section, subfigure b.}
  \label{fig:sigma_vort}
\end{figure}

\begin{figure}
  \centering
  \includegraphics[width=1\columnwidth]{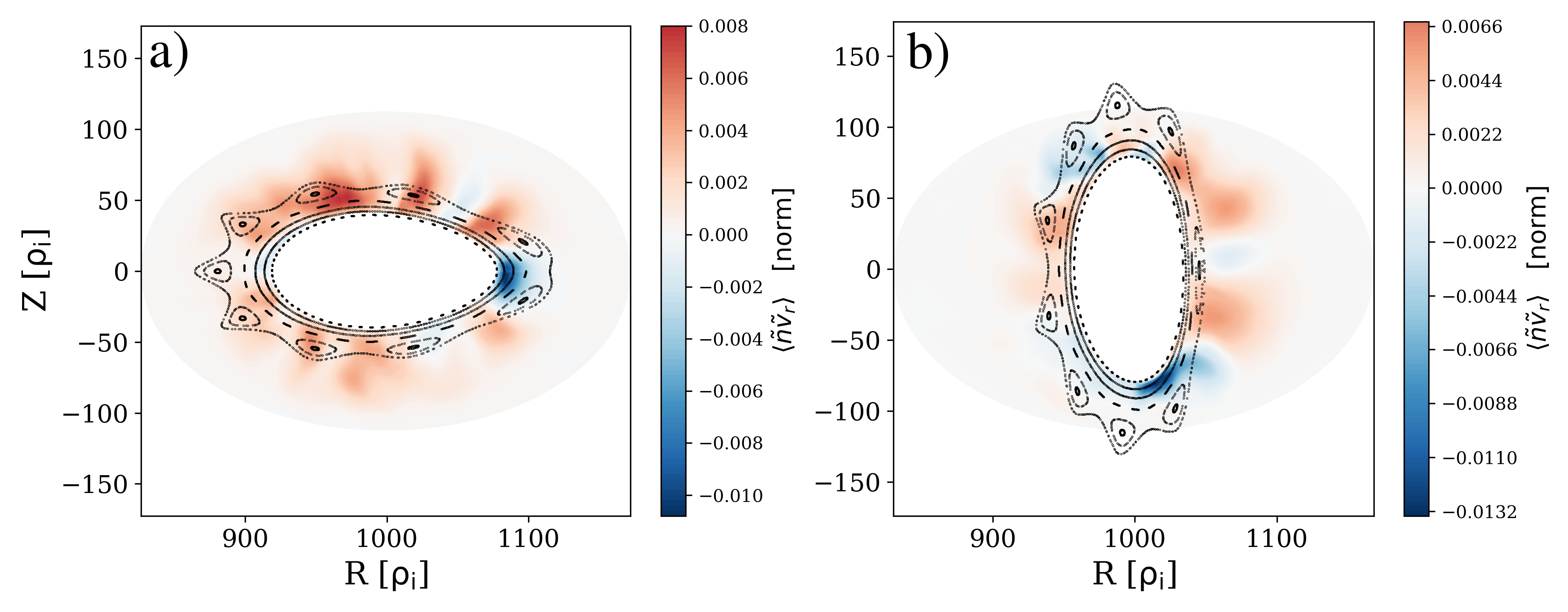}
  \caption[]{Time average of the radial perturbation flux.}
  \label{fig:nvrtilde}
\end{figure}

To determine the nature of the fluctuations, we examine central moments of the time signal, where the $n^{\mathrm{th}}$ central moment of a quantity $x$ is given by $\mu_n = \left<\left(x -\left<x\right>\right)^n\right>$ where the angled brackets ($\left<\right>$) indicate a time average. The standard deviation of $x$, $\sigma_x$, is the square root of the second moment ($n=2$) and the skewness, which will be used later, is proportional to the third moment ($n=3$) normalized to the standard deviation. The standard deviation of the density signal,~$\sigma_n$ is shown in Figure~\ref{fig:sigma_n}. 
Figure~\ref{fig:sigma_n} indicates that fluctuations are present throughout the domain. The vertical cross section, Figure~\ref{fig:sigma_n}b, indicates a decrease in $\sigma_n$ outside the outer midplane island O-point relative to the neighboring X-point regions, suggesting that fluctuations are reduced here.   
Figure~\ref{fig:sigma_n} indicates that the fluctuations, despite their large scale relative to the system size, fall off radially. 
Using the correlation coefficient, Equation~\ref{eq:xcorr}, it is determined that the fluctuations have radial correlation lengths which vary from approximately $8\rho_i$ on the outboard midplane of the vertical cross section (Figure~\ref{fig:cross_corr_diff}b) to approximately $50\rho_i$ on the inboard side of the horizontal cross section (Figure~\ref{fig:cross_corr_diff}a). Since the islands in the SOL are wider at the locations of longer correlation lengths, this poloidal variation in correlation length could be related to the island width. Furthermore, the standard deviation of the vorticity perturbations,~$\sigma_\omega$, also shows fluctuations present throughout the domain, Figure~\ref{fig:sigma_vort}. 
Figure~\ref{fig:sigma_vort}b indicates again that the outboard midplane island O-point in the vertical cross section exhibits reduced fluctuation amplitudes, whereas the neighboring X-points show stronger fluctuations. Vorticity localized to islands can in turn lead to transport around the island separatrices.  Island-localized potentials leading to flows around the island separatrix have been seen in Wendelstein {7-X}~\citep{Killer2019}. 
It is also possible to plot the radial flux of the perturbations, $\tilde{\Gamma}_\perp = \left<\tilde{n}\tilde{v}_r\right>$, illustrated in Figure~\ref{fig:nvrtilde}.
The radial perturbation flux can increase near X-points, and is indeed seen in several studies in other geometries~\citep{Choi2021, Poli2009, Hill2015}, but this phenomenon is not universal in the simulations presented here and a concrete conclusion cannot be drawn. The radial perturbation flux shown in Figure~\ref{fig:nvrtilde}b) indicates stronger outboard activity, with a change in the sign of the radial flux near the outboard midplane. 

\begin{figure}
  \centering
  \includegraphics[width=1\columnwidth]{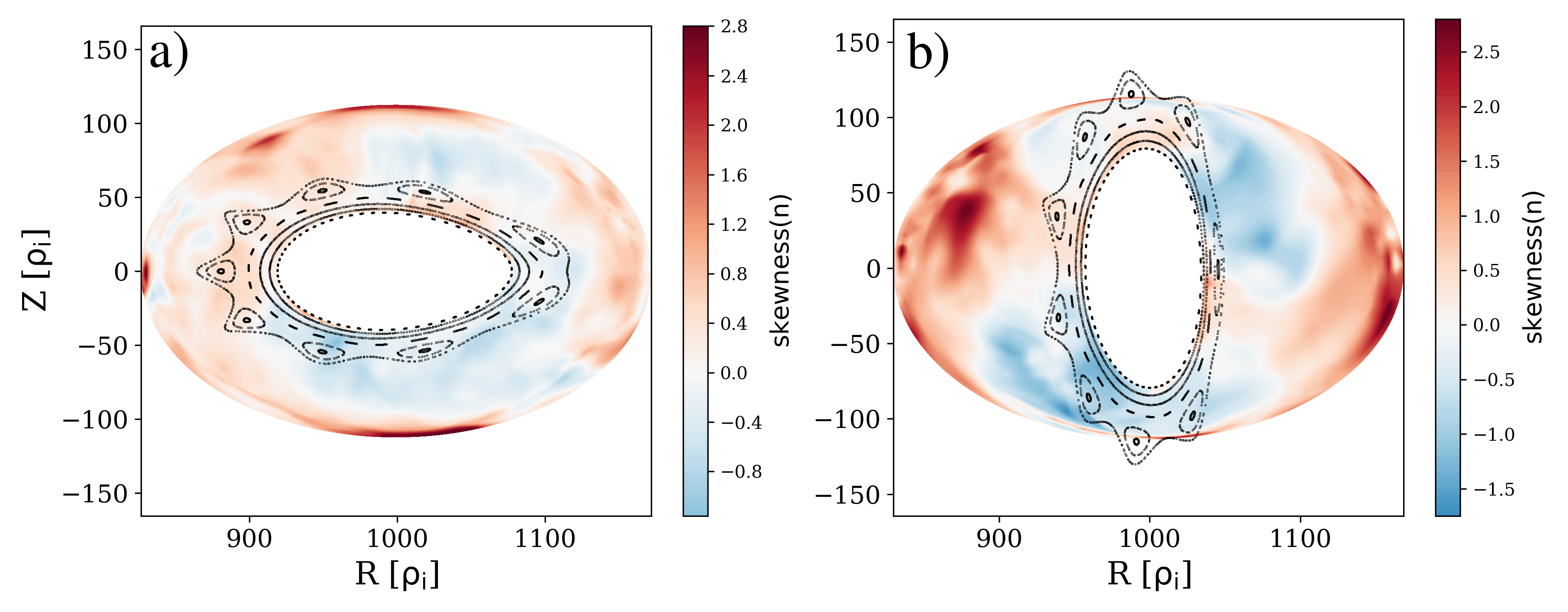}
  \caption[]{Density skewness; a positive skewness (red) indicates positive perturbations (blob-like), while negative skewness (blue) indicates the prominence of negative perturbations (holes).}
  \label{fig:skewness_n}
\end{figure}

To determine the transport nature of the fluctuations, the skewness of the profiles are plotted in Figure~\ref{fig:skewness_n}, where a positive skewness indicates blob-like transport~\citep{Walkden2017}, and negative skewness indicates areas dominated by the propagation of negative perturbations (holes).
From Figure~\ref{fig:skewness_n}, it is apparent that the transport near the outer boundary, at the inboard and outboard midplane locations mostly include positive perturbations, whereas in the middle of the domain, particularly the top and bottom of the configuration exhibit a predominantly negative skewness, where negative perturbations are more prevalent.

\section{Summary and implications}
\label{sec:discussion}
In this work, a detailed analysis of fluid turbulence in a stellarator island divertor is presented. An isothermal fluid turbulence model was used to simulate turbulence in an analytic stellarator geometry, finding that the fluctuations are present throughout the island divertor region. It was determined that the steady-state dynamics are mostly consistent with the 1/R curvature drive, but the sheath connection in the cross sections which intersect the boundary introduces a discontinuity. The plasma outside the SOL is more highly correlated with the point on the separatrix than the island O-point, indicating the separatrix as a transport channel into the private flux region. The fluctuations are exhibited throughout the domain, with a smaller amplitude near the outboard midplane island in the vertical cross section. The radial perturbation flux can, but must not necessarily manifest near island X-points. 

\subsection{Context within previous work}
\begin{figure}
  \centering
  \includegraphics[width=1\columnwidth]{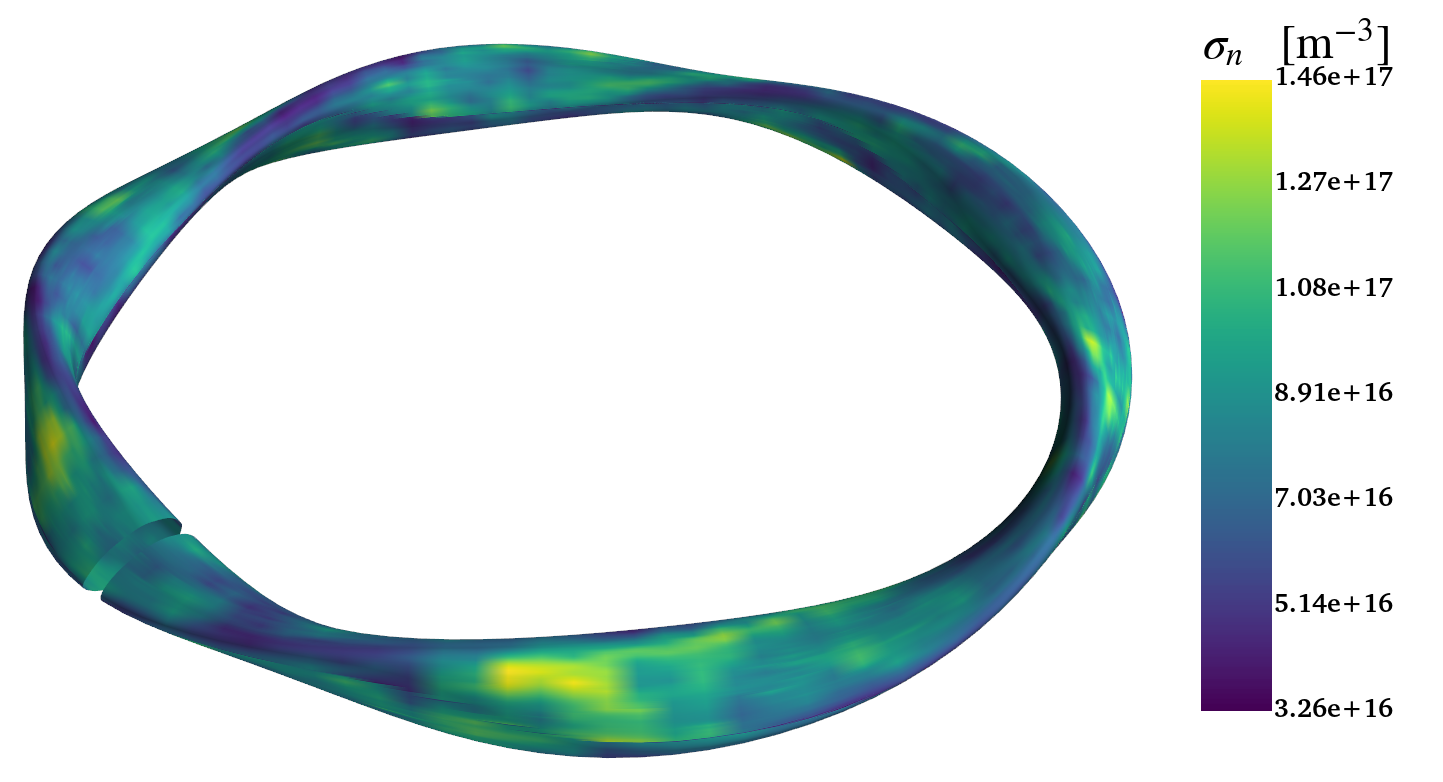}
  \caption[]{Standard deviation of the density fluctuations on the last closed flux surface; fluctuations have a higher amplitude on the outboard side.}
  \label{fig:3Dsigma}
\end{figure}

The results shown previously in~\citep{Coelho2022} suggest that despite a ballooning-dominant configuration, fluctuations are dominated by an m=4, n=5 mode and localized to the inboard side of the torus. The results presented herein indicate the fluctuation amplitudes are higher on the outboard side of the torus in the same geometry, see Figure~\ref{fig:3Dsigma}.
Furthermore, the dominant m=4 mode seen in~\citep{Coelho2022} is present but not dominant. Rather, there exists several modes -- the most prominent of which is an m=2, n=5. Generally, there does not seem to be higher-amplitude fluctuations at the inboard side. There are a few methodological differences in between this work and that shown in~\citep{Coelho2022} which could contribute to the discrepancies. Firstly, we do not simulate the core region in this work. We have chosen not to simulate the core due to the applied source at the outside of the closed-field-line region (chosen in~\citep{Coelho2022}), which could potentially influence dynamics, but future work could in principle simulate this area. Additionally, the simulation model in~\citep{Coelho2022} was nonisothermal, and therefore included other effects which cannot be captured here. 

Future work will look to relax the isothermal approximation and remove the islands from the geometry in order to more fully assess the impact of islands on the transport and perturbation dynamics, although this is a laborious task due to the non-intuitive nature of Dommaschk potentials. While it is true that by adjusting higher-order coefficients one can almost remove islands, this process is described by the author in~\citep{Dommaschk1986} as only achieved ``by trial". It could be more interesting, therefore, to explore more realistic geometries, such as the various W7-X configurations which have been more extensively investigated.

\section*{Acknowledgments}
The authors would like to thank Ant\'onio Coelho and Joaquim Loizu for many interesting discussions and providing the coefficients for the Dommaschk potential.  The efforts of the \boutxx development team have been crucial to this work. \boutxx is an open-source framework available at \url{boutproject.github.io}. 

\section*{Data availability statement}
All of the requisite files for this work are open source and can be obtained either from the lead author or at \url{github.com/bshanahan/papers}.

\section*{Funding}
This work has been carried out within the framework of the EUROfusion Consortium, funded by the European Union via the Euratom Research and Training Programme (Grant Agreement No 101052200 - EUROfusion). Views and opinions expressed are however those of the author(s) only and do not necessarily reflect those of the European Union or the European Commission. Neither the European Union nor the European Commission can be held responsible for them. 
Prepared in part by LLNL under Contract DE-AC52-07NA27344. LLNL-JRNL-849204.

\section*{Declaration of Interests}
The authors report no conflict of interest.
\section*{Author ORCID}

B Shanahan, \url{https://orcid.org/0000-0002-8766-8542}

D Bold, \url{https://orcid.org/0000-0003-0911-8606}

B Dudson, \url{https://orcid.org/0000-0002-0094-4867}
\bibliographystyle{jpp}
\bibliography{BSTING-islands}
\end{document}